\documentstyle[12pt]{article}
\begin{document}
\baselineskip .3in
\begin{center}
{\large{\bf {On Some properties of Di- Hadronic States }}} \\
\vskip .2in
B. CHAKRABARTI$^{*}$, A. BHATTACHARYA, S. MANI \\
\vskip .1in
Department of Physics, Jadavpur University \\
Calcutta 700032, India.\\
\end{center}
\vskip .3in {\centerline{\bf Abstract}} The binding energies of
di- hadronic states have been calculated assuming a 'molecular'
interaction provided by the asymptotic expression of the residual
confined gluon exchange potential between the component hadrons in
the system. Meson- meson and meson- baryon states have been
studied in detail and a mass formula has been used to calculate
total mass of the 'molecules'.

 \vskip 0.3in

 {\bf PACS No.s: 13.30.Eg, 12.40.Aa, 12.40.Qq}

 \vskip 0.3in
{\bf Keywords: Di-hadronic molecules, Tetraquarks, Pentaquarks }

\vskip .3in

*Permanent Address: Department of Physics, Jogamaya Devi College,
                    92, S P Mukherjee Road, Calcutta-700026,India
                    e- mail: $ballari<underscore>chakrabarti@yahoo.co.in$

\newpage

I. \textbf{Introduction}

\vskip 0.4in

Quantum chromodynamics allows for the existence of various multi-
quark states, and the same can also be predicted by various
phenomenological models [1-6]. While the properties of
mesons(q$\overline{q}$) and baryons(qqq) have been well documented
both in theory and experiment, the exotics ( states that do not
fit into q$\overline{q}$ or qqq states ) remain lesser known and
understood. The exotics include the tetra- quark, penta- quark and
hexa- quark states, hybrids such as $q\overline{q}g$ and gg or ggg
glueball states [7]. The recent reports on discovery of the
pentaquark $\Theta^{+}$(1540) [8-10] have revived interest in the
study of di- hadronic states in particle physics. The baryon
$\Theta^{+}$ which goes to \textit{K$^{+}$n} cannot be a 3-quark
state and its minimal quark content is uudd$\overline{s}$, making
it manifestly exotic.

Investigations into the existence of multi- quark states using the
quark and bag models [11,12] began in the early days of QCD. Non-
relativistic potential models have also been used to study these
systems [2,13]. Recently tetraquark states have been studied
extensively, specially with regard to the X(3872) particle
[14-16]. QCD spectral sum rules have been used to test the nature
of the heavy meson. The tetraquark has been studied in the
framework of the string(flux tube) model [17] where it is assumed
that di- quark- di-antiquark are connected by colour flux. Heavy
chiral unitary approach has been used to investigate interactions
between heavy vector mesons and light pseudoscalar mesons [18].
S-wave tetraquarks have also been studied using chromomagnetic
interaction with full account for flavour symmetry breaking [19].
The spectra of tetraquark and pentaquark states has been
evaluated. Complete classification has been provided in terms of
spin flavour, color and spatial degrees of freedom for tetraquarks
[20] The same has been done for 5 quark systems in terms of spin
flavour SU(6) representation [21]. However, little success has
been achieved in understanding tetra- quark and penta- quark
states as di- hadronic molecules due to the non-perturbative
nature of QCD at the hadronic scale.

The present paper seeks to examine the tetraquark and pentaquark
states as di- hadronic molecules (meson- meson and meson- baryon).
A molecular interaction of the Van der Waal's type is assumed
between the constituent hadrons [22,23]. Binding energy of the
molecules is calculated using two different wavefunctions obtained
from the Statistical model [24]. Spin hyperfine interactions have
also been taken into consideration, and a mass formula employed to
calculate total mass of the molecule. Results obtained by using
the two wavefunctions are compared to each other and to available
experimental data. The approach has been clearly outlined in
'Method', with the calculations being presented in the 'Results'
section. The work has been rounded off in 'Conclusion' which
comprises a detailed analysis of the nature of results obtained
and how well they compare to experimental data and results
obtained by other groups.

\vskip 0.4in

II. \textbf{Method}

 \vskip 0.4in
The di- hadron is constituted of either a meson- meson or a meson-
baryon held together by a Van der Waal type of interaction.  The
formula for calculating the low- lying di- hadronic molecular mass
is taken as

\begin{equation}
M_{Tot} = M_{1} + M_{2} + E_{BE} + E_{SD}
\end{equation}
where M$_{1}$ and $M_{2}$ represent the masses of the constituent
hadrons respectively, and E$_{BE}$ represents the binding energy
of the di- hadronic system and $E_{SD}$ represents the spin-
dependent term, taken separately.

E$_{BE}$ can be calculated using the formula
\begin{equation}
E_{BE}=<\Psi(r_{12})|V(r_{12})|\Psi(r_{12})>
\end{equation}
where r$_{12}$ is taken to be the radius parameter of the di-
hadronic molecule and V$(r_{12})$ is the di- hadronic molecular
potential [22, 23]. $\Psi(r_{12})$ is the wavefunction of the di-
hadronic state [24]. The wavefunctions for the ground state of the
molecule are obtained from the statistical model as
 \begin{equation}
 |\Psi(r_{12})|^{2} = \frac{315}{64\pi
 r_{0}^{9/2}}(r_{0}-r_{12})^{3/2}\theta(r_{0}-r_{12})
 \end{equation}
  and

\begin{equation}
 |\Psi(r_{12})|^{2} = \frac{8}{\pi^{2}
 r_{0}^{6}}(r_{0}^{2}-r_{12}^{2})^{3/2}\theta(r_{0}-r_{12})
 \end{equation}

 where r$_{0}$ is the radius of the molecule and
 $\theta(r_{0}-r_{12})$ represents the step function.
 The residual interaction of the confined gluon is considered
 similar to the Van der Waal's interaction and is assumed as that
 due to asymptotic expression($r_{12}\rightarrow\infty$) of the
 residual confined one gluon exchange interaction with strength
 $k_{mol}$ [22,23]. The potential is given as
 \begin{equation}
 V(r_{12}) = \frac{-k_{mol}}{r_{12}}e^{-C^{2}r_{12}^{2}/2}
 \end{equation}
 where $k_{mol}$ is the residual strength of the strong
 interaction molecular coupling and C is the effective colour
 screening of the confined gluons.
 The radius $r_{0}$ of the di- hadronic molecule in Eqs. 3 and 4 is obtained by employing an
 additive rule to the radii of constituent hadrons. We take $r_{o}$ =  $r_{1}$ +
 $r_{2}$, where r$_{1}$ and r$_{2}$ represent the individual radii of the
 hadrons constituting the molecule, respectively.

 Now, using Eqs. 2, 3 and 5 we get an integral for $E_{BE}$ which
 yields

\begin{equation}
E_{BE}=
\frac{315k_{mol}}{16r_{0}}0.114286\textit{Hyp}PFQ[(3/2,1),(2.75,2.25),-\beta]
if Re\beta>0
\end{equation}

where Hyp PFQ is a hypergeometric function. Now, using Eqs. 2,4
and 5 we get

\begin{equation}
E_{BE}= \frac{32k_{mol}}{\pi
r_{0}}[\frac{-0.75+0.5\beta}{\beta^{2}} +
\frac{e^{-\beta}}{\beta^{5/2}} +
\frac{0.66467e^{-\beta}Erfi[\surd\beta]}{\beta^{5/2}}]
\end{equation}

where Erfi[] represents an error function and $\beta$=
$\frac{C^{2}r_{12}^{2}}{2}$ and C=50 MeV [25] has been
substituted. The experimental mass for $\Theta^{+}$ is 1540 MeV
[8] being known, the values of masses of the constituent hadrons
($K^{+}$ and n) are substituted in Eq.1 to find the corresponding
binding energy. This is then used in Eq.7 to calculate the value
of k$_{mol}$. It is found k$_{mol}$=0.47, which is fixed as the
strength of the residual confined gluonic interaction
corresponding to the di- hadronic wavefunction in Eq.3. Working
the same way, $k_{mol}$=0.52 is calculated for the wavefunction in
Eq. 4. The radii considered are $r(\pi)=5.38 GeV^{-1}$, $r(K)=4.77
GeV^{-1}$, $r(\rho)=4.75 GeV^{-1}$ [26], $r(D_{s})=4.3 GeV^{-1}$
[27], $r(\omega)= 4.765 GeV^{-1}$, $r(\phi)=5.00 GeV^{-1}$
[28].Similarly $r(p)=6 GeV^{-1}$, $R(n)=4.7 GeV^{-1}$[29],
$r(\Lambda_{c}^{+})=5.727 GeV^{-1}$, $r(\Sigma_{c}^{+})=3.386
GeV^{-1}$, $r(\Xi_{c}^{0})=2.404 GeV^{-1}$ [30], $r(\Sigma)=3.9
GeV^{-1}$ [31]. Experimental values of the respective meson and
baryon masses have been considered for calculation [32].

$E_{SD}$, the contribution to total mass from spin hyperfine
interaction is given by the following expression [33]
\begin{equation}
E_{SD}=\frac{8}{9} \frac{\alpha_{s}}{m_{1}m_{2}}
\overrightarrow{S_{1}}.\overrightarrow{S_{2}}|\Psi(0)|^{2}
\end{equation}

where $m_{1}$ and $m_{2}$ are the individual masses of the
constituent hadrons in the di hadronic molecule,$\alpha_{s}$ is
the strong interaction constant and $S_{1}$ and $S_{2}$ are the
spins of the hadrons involved. $|\Psi(0)|^{2}$ is the value of the
di- hadronic wavefunction at the origin. The spin contribution on
calculation is found to be extremely small in comparison with
binding energy of the molecules. The values of binding energy and
spin- dependent term and total mass calculated for the di-
hadronic molecules are displayed in TableI (Tetra- quark states)
and Table II(Penta- quark states)and Table III (hexa-quark
states). Available experimental data are presented alongside in
the tables to provide a ready comparison with obtained values.

\vskip 0.4in

IV. \textbf{Conclusion}

 \vskip 0.4in

The binding energies and masses of exotic states (tetraquark and
pentaquark) have been calculated by considering them as
di-hadronic molecules. The values are calculated using two
separate wavefunctions (Eq. 3 and Eq. 4) obtained using linear and
harmonic potentials respectively. The values obtained in both
cases are found to be in close agreement with no serious
discrepancy. This may be considered to suggest that for any
wavefunction arising from a smooth background potential, the
calculated values will be independent of the  actual potential.
The experimental states are exotics whose spin- parity do not
match with the expected quark- antiquark structure for mesons or
the 3- quark structure for baryons, Hence pseudoscalar-
pseudoscalar di mesonic states and vector- vector di mesonic
states are assigned parity charge- conjugation PC '++' and
pseudoscalar- vector di mesonic states have PC '+-'. The values of
$J^{PC}$ have also been indicated in the tables and those
experimental states whose $J^{PC}$ values match the predicted
values have been considered for comparison. In case of di-mesonic
states, a number of experimentally observed states were available,
to which the calculated values have been compared, wherever
possible. The calculated results are found to be in good agreement
with experiment. The $\pi-K$ state is found to have a mass of
approximately 0.73 GeV which is nearly the same as that calculated
by Rai et al($\sim0.72 GeV$) [25]. The binding energies of tetra-
quark states are found to lie within a range of 0.094 GeV to 0.114
GeV, whilst the masses range from 0.373 to 2.575 GeV. The low-
lying pentaquark states have been treated as meson- baryon
molecules. The experimentally observed $\Theta^{+}$ state[8] has
been taken as reference for calculating the value of the
interaction strength parameter. The obtained values have been
compared to the few available experimental data and are found to
be in favourable agreement. The calculated total mass for the
$\Sigma-K$ state (1.77 GeV) is found to be close to the estimate
given by Rai et al [25]. The well known $\bigwedge$ particle
(1.405 GeV)[34] is found to almost exactly match the predicted
$\pi-\Sigma$ molecule (1.41GeV). The binding energies of meson-
baryon molecules are found to range from 0.0848 to 0.145 GeV
whilst the total mass ranges from 1.41 to 3.3 GeV. In calculating
the binding energies and total mass of hexa- quark particles
experimental results were not available for comparison. Despite
theory suggesting that the hexa- quark state should be more stable
than the penta- quark state, experimental evidence has been found
in support of the latter, not the former. Binding energies of the
hexa- quark states are found to range from 0.0938GeV to 0.133GeV,
and mass varies from 2.012GeV to 2.479GeV.

We have recently successfully predicted the mass of the pentaquark
$\Theta^{+}$ using a diquark approach[35]. This suggests that the
same approach may be extended to studying other multiquark states
as has been done here and we look forward to doing the same in our
future works. The good agreement of the theoretically calculated
results with available experimental data and other theoretical
works encourages us to suggest that a number of the predicted
states may be experimentally detected in the near future.

\newpage
V. \textbf{Results}

\vskip 0.4in

 TableI: Binding Energies and Masses of Di-Meson States

\vskip 0.4in

\begin{tabular}{|r|r|r r|r r|r|r|}
\hline
 $  $&  &$Linear$&$Potential$&$Harmonic$&$Potential$&$  $\\

 $States$&$J^{PC}$&$E_{BE}+E_{SD}$&$M $&$E_{BE}+E_{SD}$&$M$&$M_{Exp}$\\
 $  $&  &$(GeV)$&$(GeV)$&$(GeV)$&$(GeV)$&$(GeV)$\\
\hline

 $\pi-\pi$&$0^{++}$&0.0948&0.3738&0.0944&0.3734&0.4-1.2\\
 $\pi-K$&$0^{++}$&0.1010&0.7335&0.1005&0.7355&---\\
 $\pi-\omega$&$1^{+-}$&0.101&1.023&0.1005&1.0227&---\\
 $\pi-\phi$&$1^{+-}$&0.0985&1.257&0.0978&1.256&1.235 $(b_{1})$\\
 $\pi-D_{s}$&$0^{++}$&0.1059&2.213&0.1055&2.212&2.317 $(D_{s0}^{*}$\\
 $\pi-\rho$&$1^{+-}$&0.0985&1.257&0.0979&1.256&1.17 $(h_{1})$\\
 $K-\omega$&$1^{+-}$&0.1075&1.3839&0.1074&1.3838&1.4 $(K_{1})$\\
 $K-\phi$&$1^{+-}$&0.1049&1.618&0.1045&1.6167&---\\
 $K-D_{s}$&$0^{++}$&0.1142&2.575&0.1132&2.575&---\\
 $K-K$&$0^{++}$&0.1075&1.0948&0.1074&1.0947&1.27 $(K_{1})$\\
 $K-\rho$&$1^{+-}$&0.1076&1.376&0.1076&1.376&1.41 $(K_{1)}$\\
 $\rho-\rho$&$1^{++}$&0.108&1.659&0.1078&1.658&1.69 $(\rho_{3})$\\
 $\rho-\omega$&$2^{++}$&0.1077&1.665&0.1077&1.665&---\\
 $\rho-\phi$&$2^{++}$&0.1052&1.9&0.1047&1.898&---\\

 $\phi-\phi$&$2^{++}$&0.1024&2.1144&0.102&2.14&2.01 $(f_{2})$\\
\hline
\end{tabular}

\newpage
\vskip 0.4in

TableII: Binding Energies and Masses of Di-Baryon States

\vskip 0.4 in

\begin{tabular}{|r|r|r r|r r|r|r|}
\hline
 $  $&  &$Linear$&$Potential$&$Harmonic$&$Potential$&$  $\\

 $States$&$J^{P}$&$E_{BE}+E_{SD}$&$M $&$E_{BE}+E_{SD}$&$M$&$M_{Exp}$\\
 $  $&  &$(GeV)$&$(GeV)$&$(GeV)$&$(GeV)$&$(GeV)$\\
\hline

 $\pi-N$&$1/2^{+}$&0.1017&1.1807&0.1014&1.1804&1.23$(P_{33})$\\
 $\pi-P$&$1/2^{+}$&0.089&1.1672&0.0889&1.1667&---\\
 $\pi-\Sigma$&$1/2^{+}$&0.0855&1.4118&0.0848&1.4168&1.4$(S_{01})$\\
 $\pi-\Sigma_{c^{+}}$&$1/2^{+}$&0.1195&2.7125&0.1195&2.7125&---\\
 $\pi-\Lambda_{c}^{+}$&$1/2^{+}$&0.0918&2.5104&0.9136&2.5170&2.52$(\Sigma_{c})$\\
 $\pi-\Xi_{c}^{0}$&$3/2^{+}$&0.1345&2.741&0.1332&2.742&2.645$(\Xi_{c})$\\
 $K-P$&$1/2^{-}$&0.0947&1.5258&0.093&1.510&---\\
 $K-\Sigma$&$1/2^{-}$&0.0904&1.77&0.0899&1.77&1.75($\Sigma$)\\
 $K-\Sigma_{c}^{+}$&$1/2^{-}$&0.0973&2.877&0.0973&2.877&---\\
 $K-\Xi_{c}^{0}$&$1/2^{-}$&0.145&3.109&0.144&3.054&---\\
 $\rho-N$&$1/2^{+},3/2^{+}$&0.1088&1.826&0.1086&1.823&1.9$(P_{35)})$\\
 $\rho-P$&$1/2^{+},3/2^{+}$&0.0948&1.8078&0.0945&1.8079&---\\
 $\rho-\Sigma$&$1/2^{+}$&0.0898&2.0508&0.0902&2.055&---\\
 $\rho-\Sigma_{c}^{+}$&$1/2^{+}$&0.1289&3.3526&0.1289&3.352&---\\
 $\omega-N$&$1/2^{+},3/2^{+}$&0.1088&1.83&0.108&1.83&---\\
 $\omega-\Sigma$&$1/2^{+}$&0.0898&2.06&0.089&2.06&---\\
 $\omega-P$&$1/2^{+},3/2^{+}$&0948&1.814&0.095&1.814&---\\
 $\phi-N$&$1/2^{+},3/2^{+}$&0.1055&2.064&0.1058&2.064&---\\
 $\phi-P$&$1/2^{+},3/2^{+}$&0.927&2.05&0.0928&2.05&---\\
 $\phi-\Sigma$&$1/2^{+}$&0.088&2.301&0.089&2.301&---\\
 \hline
 \end{tabular}
 \vskip 0.4in

 TableIII: Binding Energies and Masses of Di-Baryon States

\vskip 0.4in
\begin{tabular}{|r|r|r r|r r|r|r|}
\hline
 $  $&  &$Linear$&$Potential$&$Harmonic$&$Potential$&$  $\\

 $States$&$J^{PC}$&$E_{BE}+E_{SD}$&$M $&$E_{BE}+E_{SD}$&$M$&$M_{Exp}$\\
 $  $&  &$(GeV)$&$(GeV)$&$(GeV)$&$(GeV)$&$(GeV)$\\
\hline

 $N-N$&&0.133&2.012&0.1297&2.008&---\\
 $N-\Sigma$&&0.1106&2.428&0.109&2.241&---\\
 $P-\Sigma$&&0.0985&2.229&0.0976&2.229&---\\
 $\Sigma-\Sigma$&&0.0938&2.479&0.0939&2.479&---\\
 \hline

\end{tabular}

\newpage
VI. \textbf{References}

\vskip .4in

\noindent [1]. J L Rosner; Phys.Rep. {\bf 11 }, 89 (1974).

\noindent [2]. T Barnes, K Dooley, N Isgur; Phys.Rev.Lett. {\bf B
183}, 210 (1987).

\noindent [3]. R L Jaffe;  Phys.Rev.  {\bf D 38}, 195 (1977).

\noindent [4]. S Fleck, C Grignoux, J Richard; Phys. Lett. {\bf B
220}, 616 (1989).

\noindent [5]. S Zouzou et al; Z. Phys. {\bf C 30 }, 457 (1986).

\noindent [6]. H J Lipkin; Phys. Lett. {\bf B 172}, 242 (1986).

\noindent [7]. PDG, C Cass et al. ; Eur. Phys. J{\bf C 3}, 1
(1998), K Hagiwara et al. ; PDG Phys. Rev. {\bf D 66}, 1 (2002),
Erich Braaten, Masaki Kusonski, Shmuel Nussinov;
arXiv:hep-ph/0404161  v1 (2004).

\noindent [8]. T Nakano et al. ; Phys. Rev. Lett.  {\bf 91},
012002 (2003).

\noindent [9]. V V Barwin et al. [DIANA Collaboration];
arXiv:hep-ex/ 0307018.

\noindent [10]. S Stepanyan [CLAS Collaboration]; arXiv:
hep-ex/0307018.

\noindent [11]. R L Jaffe; Phys. Rev. {\bf D 15}, 281(1977).

\noindent [12]. D Strottman; Phys. Rev. {\bf D 20}, 748 (1979).

\noindent [13]. J Weinstein, N Isgur; Phys. Rev.
 {\bf D 27} 588 (1983).

\noindent [14]. R D Mattheus, S Narison, M Nielsen, J M Richard;
Phys. Rev. {\bf D }, ----

 \noindent [15]. Muneyuki Ishida, Shin Ishida, Tomahito Maeada;
 arXiv:hep=ph/0509212 (2005).
\noindent [16]. Eric S Swanson; Phys. Rep. {\bf 429}, 243 (2006)

 \noindent [17]. Mashaharu Iwasaki, Takahiko Fukutome; Phys. Rev. {\bf D 72}, 094016 (2005).

\noindent [18]. Feng Kun Guo, Peng- Nian Shen, Huan Ching Chiang;
Phys. Lett. {\bf B 647}, 133 (2007).

\noindent [19]. Hogasen Buccella, Richard et al. ;Eur. Phys. J.
{\bf C 49},743 (2007).

\noindent [20]. E Santopinto, Galata Guiseppe; Phys Rev. {\bf C 75
4},04 (2007)

\noindent [21]. R Bijker, M M Giannini, E Santopinto; Eur. Phys. J
{\bf A 22}, 319 (2004).

\noindent [22]. P C Vinodkumar, J N Pandya, S B Khadkikar;
Pramana-J Phys.{\bf  39},47 (1992).

\noindent [23]. P C Vinodkumar, J N Pandya, Ajay Kumar Rai; DAE-
BRNS Symposium on Nucl. Phys. {\bf  B}, 334 (2003).

\noindent [24].A Bhattacharya  et al, Eur. Phys. J  {\bf C 2}, 671
(1998).

\noindent [25].  Ajay Kumar Rai, J N Pandya, P C Vinodkumar;
Indian J. Phys.{\bf 80(4)}, 387 (2006).

\noindent [26].  S N Banerjee et al; Int. J. Mod. Phys. {\bf A 2},
1829 (1987)

\noindent [27].  B Chakrabarti; Mod. Phys. Lett.{\bf A 12 No 28},
2133 (1997).

\noindent [28].  S N Banerjee et al; Int. J. Mod. Phys. {\bf A 4},
943 (1989).

\noindent [29].  S N Banerjee et al; Can. J. Phys. {\bf 66}, 749,
(1988).

\noindent [30].  B Chakrabarti et al; Phys. Scr. {\bf 61}, 49,
(2000).

\noindent [31].  arXiv:hep-ex/010-6053v2.

\noindent [32].  W M Yao et al; J. Phys. G: Nucl. Part. Phys. {\bf
33}, 1 (2006).

\noindent [33].  Ajay Kumar Rai, J N Pandya, P C Vinodkumar; Nucl.
Phys. {\bf  A782},406c-409c  (2007).

\noindent [34].  M Lutz et al; Nucl. Phys. {\bf A700}, 193,
(2002); J Nacher et al; Phys. Lett. {\bf B455}, 55, (1999)

\noindent [35].  B. Chakrabarti et al; Nucl. Phys. {\bf A782/1-4},
392c, (2007).
\end{document}